\documentclass[conference]{IEEEtran}
\IEEEoverridecommandlockouts
\usepackage{cite}

\usepackage{algorithmic}
\usepackage{graphicx}
\usepackage{mathtools, nccmath}

\usepackage{xcolor}
\usepackage{amsmath,amssymb,amsthm}

\usepackage{multirow}
\usepackage{lettrine}

\begin{document}

\title{Quantum Annealing for Automated Feature Selection in Stress Detection}

\author{\IEEEauthorblockN{ Rajdeep Kumar Nath$^{1}$, Himanshu Thapliyal$^{1*}$ and Travis S. Humble$^{2}$}
\IEEEauthorblockA{\textit{$^{1}$Department of Electrical and Computer Engineering, University of Kentucky, Lexington, KY, USA} \\
\IEEEauthorblockA{\textit{$^{2}$Oak Ridge National Laboratory, Oak Ridge, TN, USA} \\
*Corresponding Author: hthapliyal@ieee.org
}}

}

\maketitle

\begin{abstract}
We present a novel methodology for automated feature subset selection from a pool of physiological signals using Quantum Annealing (QA). As a case study, we will investigate the effectiveness of QA-based feature selection techniques in selecting the optimal feature subset for stress detection. Features are extracted from four signal sources: foot EDA, hand EDA, ECG, and respiration. The proposed method embeds the feature variables extracted from the physiological signals in a binary quadratic model. The bias of the feature variable is calculated using the Pearson correlation coefficient between the feature variable and the target variable.  The weight of the edge connecting the two feature variables is calculated using the Pearson correlation coefficient between two feature variables in the binary quadratic model. Subsequently, D-Wave's clique sampler is used to sample cliques from the binary quadratic model. The underlying solution is then re-sampled to obtain multiple good solutions and the clique with the lowest energy is returned as the optimal solution. The proposed method is compared with commonly used feature selection techniques for stress detection. Results indicate that QA-based feature subset selection performed equally as that of classical techniques. However, under data uncertainty conditions such as limited training data, the performance of quantum annealing for selecting optimum features remained unaffected, whereas a significant decrease in performance is observed with classical feature selection techniques. Preliminary results show the promise of quantum annealing in optimizing the training phase of a machine learning classifier, especially under data uncertainty conditions.  

\end{abstract}

\section{Introduction}
\begin{figure}[h]
\centering
\includegraphics[trim= 0cm 0cm 0cm 0cm, scale=0.25]{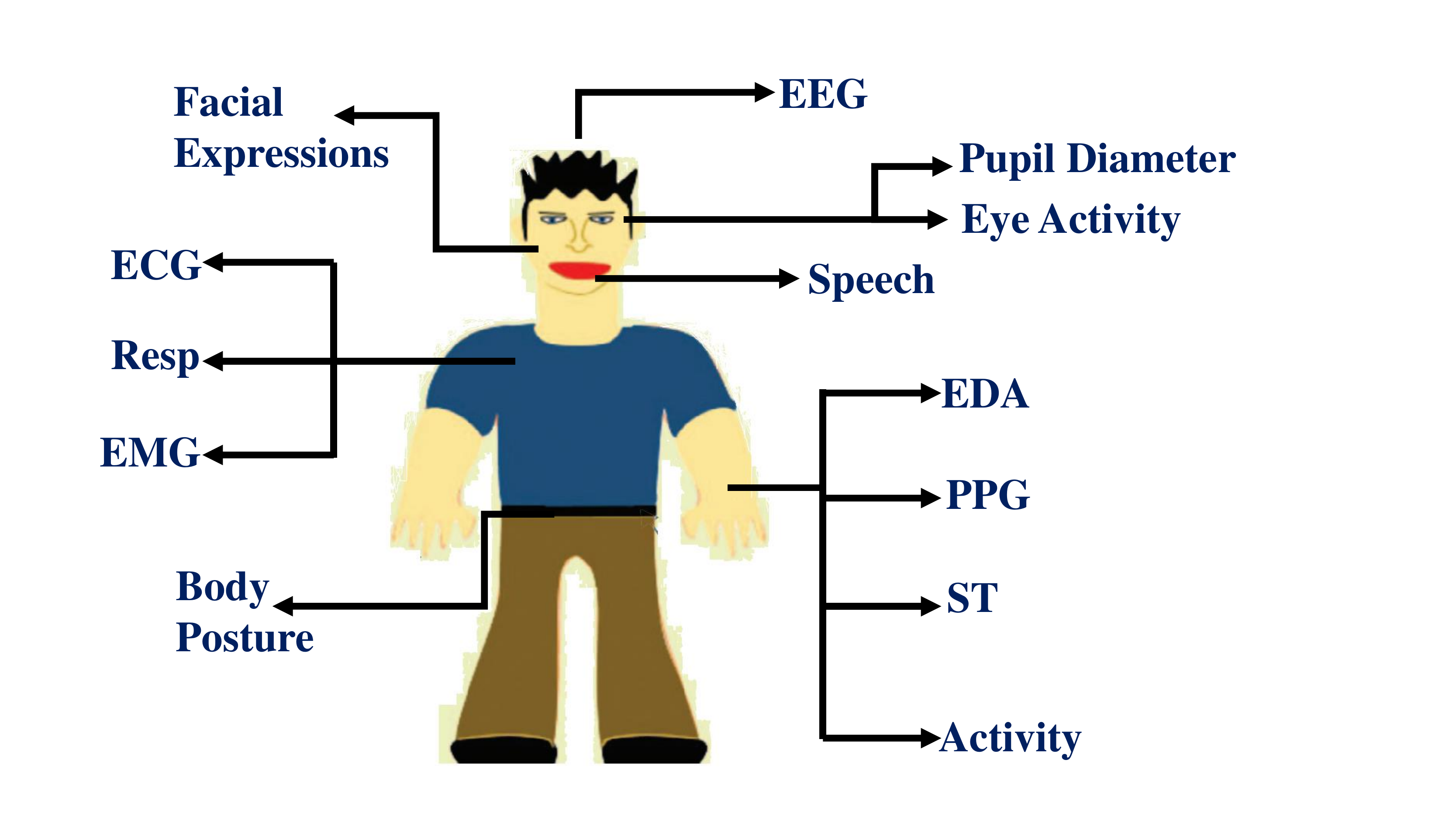}
\caption{Some of the various physiological measures used for stress detection.}
\label{physiological_sources}
\end{figure}

Stress detection is an important research topic in health informatics. Experiencing frequent stress can be a major health concern that can pose a threat to the physical and mental stability of an individual in the long run. Chronic effects of stress can cause physiological abnormalities such as hypertension, stroke, obesity, and diabetes, and psychological conditions such as cognitive impairments which might lead to the development of Alzheimer’s Disease in older adults \cite{001}\cite{002}. Although stress is an unavoidable aspect of our daily lives, monitoring and managing stress can significantly reduce the long-term negative effects of stress. Monitoring and quantifying stress reliably is not a trivial task as stress is a complex phenomenon and is influenced by several factors such as environmental factors, mental workload, task-specific stressors, etc. Researchers have adopted various methods for detecting stress. The most popular methods for detecting stress are through monitoring of physiological signals such as EDA (Electrodermal Activity), ECG (Electrocardiogram), PPG (Photoplethysmogram), RESP (Respiration) and EEG (Electroencephalogram). Together with these, researchers have also analyzed behavioral and activity-based features for stress detection \cite{003}\cite{004}. Further, other physiological measures such as eye activity, pupil diameter, speech, etc. have also been used for stress detection \cite{007}. Figure \ref{physiological_sources} shows some of the various measures adopted for monitoring stress levels. 

Although studies have been found that combining different physiological measures usually increases the performance of the stress classification model, this might increase the complexity of the system. Moreover, the usefulness of a particular physiological measure in characterizing stress depends on the application \cite{005}. For example, it has been found that EDA-based features are better in detecting stressed states during tasks that involve physical activity \cite{006}. Similarly, it is found that ECG features are sensitive to changes in the context of physical activity \cite{006}. Hence, identifying the optimal set of physiological parameters that can most accurately correlate the changes in the physiological response with stress is an important research topic \cite{005}.

Usually feature selection methods are used for reducing the set of extracted features before training a machine learning classifier. Once the important set of features are identified in the training phase, subsequently, we will only need to extract those features for prediction in the deployment phase. Typically, features are selected by ranking the features based on their association with the target variable. This association is usually quantified by some statistical measures such as correlation coefficient, mutual information, p-value, etc. Subsequently, the best set of features is selected by following a greedy search technique \cite{008}\cite{003}. However, a major problem with greedy search is that it is not guaranteed to reach the global minimum of the solution space because it does not evaluate all the possible combinations. To evaluate all the possible combinations, an exhaustive greedy search is required which is computationally very expensive. For example, if we want to choose the best performing 20 features from a set of 50 features, there will be  approximately $4.7 \times 10^{13}$ combinations. 

In this work, we explore the application of quantum annealing for selecting the optimal set of features from EDA signals collected from hand and foot, ECG, and RESP signals. The EDA, ECG, and RESP signals were downloaded from a publicly available database \cite{010}. These physiological signals were collected during a driving task. During the driving task, the driver was required to start from rest and then drive through a series of city and highway driving segments. The physiological signals were annotated with "Low", "Medium", and "High" stress for driving segments during rest, highway, and city respectively. The quantum annealing algorithm is implemented using a physical quantum computer realized by D-Wave systems \cite{009}. 

The contribution of this work is as follows:
\begin{itemize}
\item A hybrid classical-quantum machine learning pipeline is proposed for optimizing the number of features used for training the machine learning classifier.
\item The proposed hybrid classical-quantum machine learning pipeline is used to solve the problem of selecting an optimum subset of features from a pool of physiological signals for detecting stress in automobile drivers.
\item Performance of the QA-based feature selection technique is compared with typically adopted classical feature selection techniques for stress detection under limited training data.
\end{itemize} 

The paper is organized as follows: Section \ref{background} will discuss the background on quantum annealing briefly. Section \ref{proposed_method} will outline the proposed classical-quantum hybrid training pipeline. Section \ref{result} will discuss the results, and finally Section \ref{conc} will conclude the work. 

\section{Background on Quantum Annealing and Quantum Processing Unit}
\label{background}
Quantum annealing is a noisy variant of the adiabatic model of quantum computation that allows relaxation on the strict requirements of the ideal adiabatic condition. It is a heuristic search technique to find the global minimum of an objective function. Theoretically, it has been proven that quantum annealing is always guaranteed to reach the global minimum with a faster rate of convergence than classical annealing \cite{011}. In greedy search techniques, the solution can get stuck in local minima. Quantum annealing escapes local minima by tunneling through the long energy barriers.

In D-Wave's quantum annealer, the quantum computation is performed by the QPU (Quantum Processing Unit). The QPU is mainly a collection of qubits which are arranged in graph structures. Some of the qubits are interconnected to each other through couplers. While solving an optimization problem, the qubits are initialized with the bias of the variables that needs to be optimized. The strength of the couplers connecting any two qubits are initialized with the weight of the connection between the two variables. After the annealing the final values of the biases and weights are returned as the solution to the optimization problem. The optimized solutions are those that have the lowest energy configuration. The underlying solutions can be then be re-sampled n-times to obtain different solutions to the same problem. 

\section{Proposed Work}
\label{proposed_method}
\begin{figure*}[h]
\centering
\includegraphics[trim= 0cm 0cm 0cm 0cm, scale=0.5]{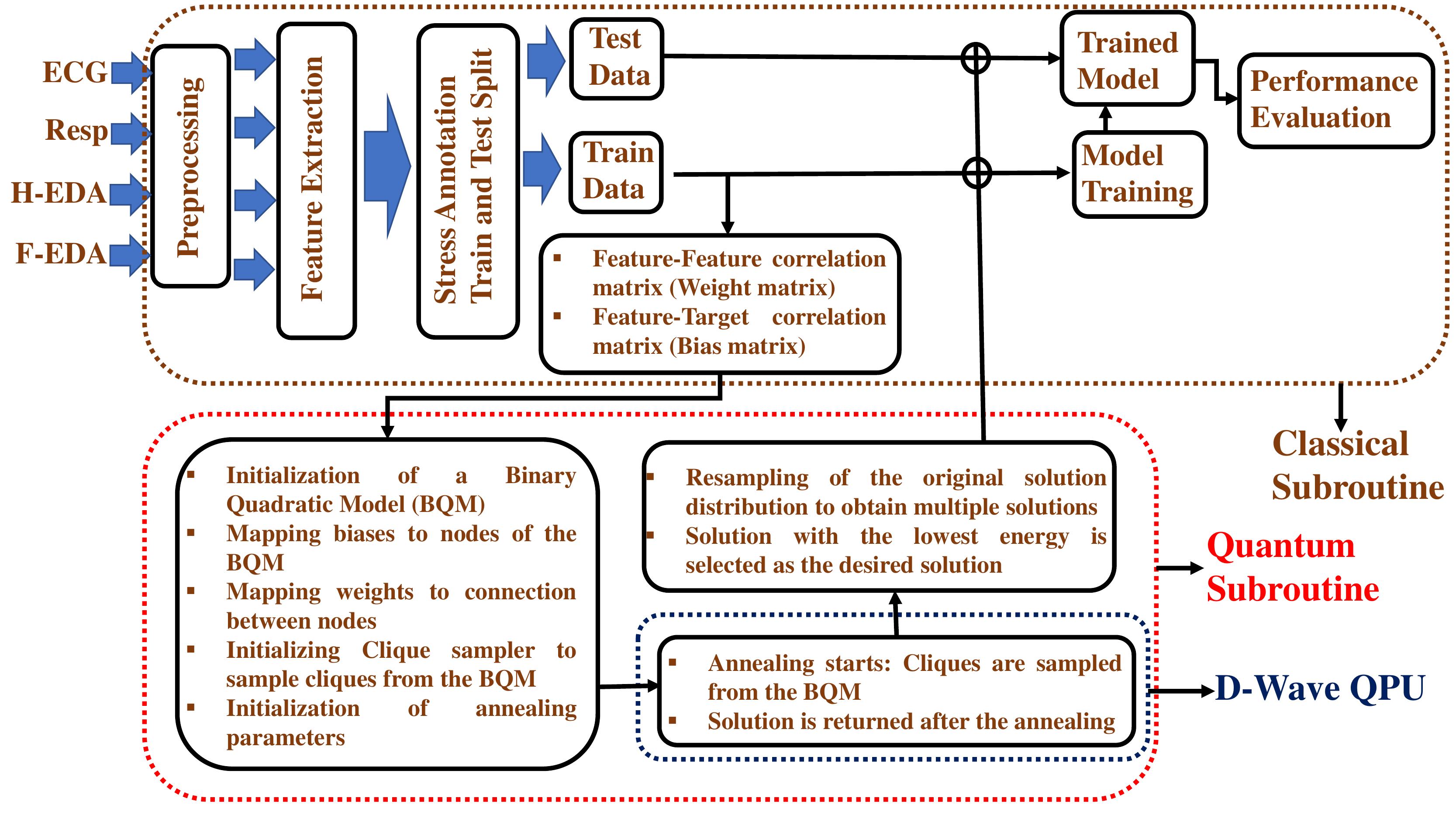}
\caption{Overview of the proposed work.}
\label{overview}
\end{figure*}

Figure \ref{overview} shows the design of the proposed hybrid classical-quantum machine learning pipeline. This hybrid pipeline consists of a classical subroutine and a quantum subroutine. We will discuss these subroutines in detail in this section.

\subsection{Classical Subroutine}
The classical subroutine executes the preprocessing and feature extraction stages of the machine learning pipeline. 
\subsubsection{Preprocessing}
In this module, the EDA signals from hand and foot, ECG, and RESP signals are normalized and filtered using a low-pass Butterworth filter of order 5 and the signal components greater than 1 Hz for EDA, 40 Hz for ECG, and 10 Hz for Resp signals were cut off.  
\subsubsection{Feature Extraction}
Features were extracted using a running window of 100 seconds and an overlap of 50 seconds for all four physiological signals. Six features each were extracted from the EDA signal collected from hand and foot. These six features comprised of the mean, variance, number of peaks in a time window, sum of the EDA peaks, and the sum of the response duration of the EDA signal.

For the ECG signal, a total of 15 time-domain features and 6 frequency domain features were extracted. The time-domain features comprised of the various statistical measure of the R peaks of ECG signal and the mean, maximum, minimum, and standard deviation of the heart rate extracted from the ECG signal. The frequency-domain features comprised of low power, high power, and very low and high power components of ECG signals, along with their ratios. 

For the RESP signal, a total of 6 features were extracted, which included the mean and variance of the respiration signal. The remaining four features were the average power in the four frequency bands, (0-0.1 Hz), (0.1-0.2 Hz), (0.2-0.3 Hz), and (0.3-0.4 Hz). More details on the feature extraction from these signals can be found in \cite{010}.
\subsubsection{Train and Test Split}
The entire feature set is first annotated with their respective stress labels, that is low stress, medium stress, and high stress. Subsequently, the feature set is divided into train and test sets in the approximate ratio of 70-30. The train set is then used to compute the bias and weight matrix by computing the pair-wise feature-target correlation, and feature-feature correlation respectively. In this work, we have used the Person correlation coefficient to quantify the feature-feature and feature-target correlation. It is to be noted that only the weight and the bias matrices are offloaded to the quantum subroutine and not the entire training dataset.    

\subsection{Quantum Subroutine}
The quantum subroutine is comprised of a classical software interface that maps problem instances to the D-Wave QPU. The quantum subroutine consists of a software interface that translates the real-world problem instance into a representation that can be solved by the D-Wave's QPU. To translate the problem, a binary quadratic model is initialized. The nodes represent the features, and the strength of the connection between the nodes represents the interaction between the feature variables. The values of the nodes are initialized with the biases obtained from the feature-target bias matrix from the training phase, and the strength of the connection between the nodes is initialized with the feature-feature weight matrix from the training phase. 

After the initialization of the weights and biases, D-Wave's clique sampler is used to sample cliques with the lowest energy configuration. The solution returned by the D-Wave clique sampler is resampled multiple times to obtain multiple close-to-good solutions and the clique with the lowest energy is chosen as the best solution. The nodes that form the cliques with the lowest energy configuration is hypothesized as the best performing optimum subset of features. 

\section{Results and Analysis}
\label{result}

In this section, we will discuss the performance of the QA-based feature selection technique with typically adopted feature selection techniques for stress detection such as Pearson correlation ranking-based feature selection, p-value based feature selection, and mutual-information-based feature selection. For simplicity, the default parameters of the quantum annealer was used in this work. 

\subsection{Evaluation Objective}
The objective of our analysis is to verify the effectiveness of the QA-based feature selection algorithm in selecting the optimum set of features for stress level prediction. In this section, we will perform a quantitative and qualitative evaluation of the QA-based feature selection method. A comparison with classical feature selection techniques under 100\%, 30\%, 20\%, and 10\% of the training data will also be analyzed.

\begin{figure*}[h]
\centering
\includegraphics[trim= 0cm 0cm 0cm 0cm, scale=0.42]{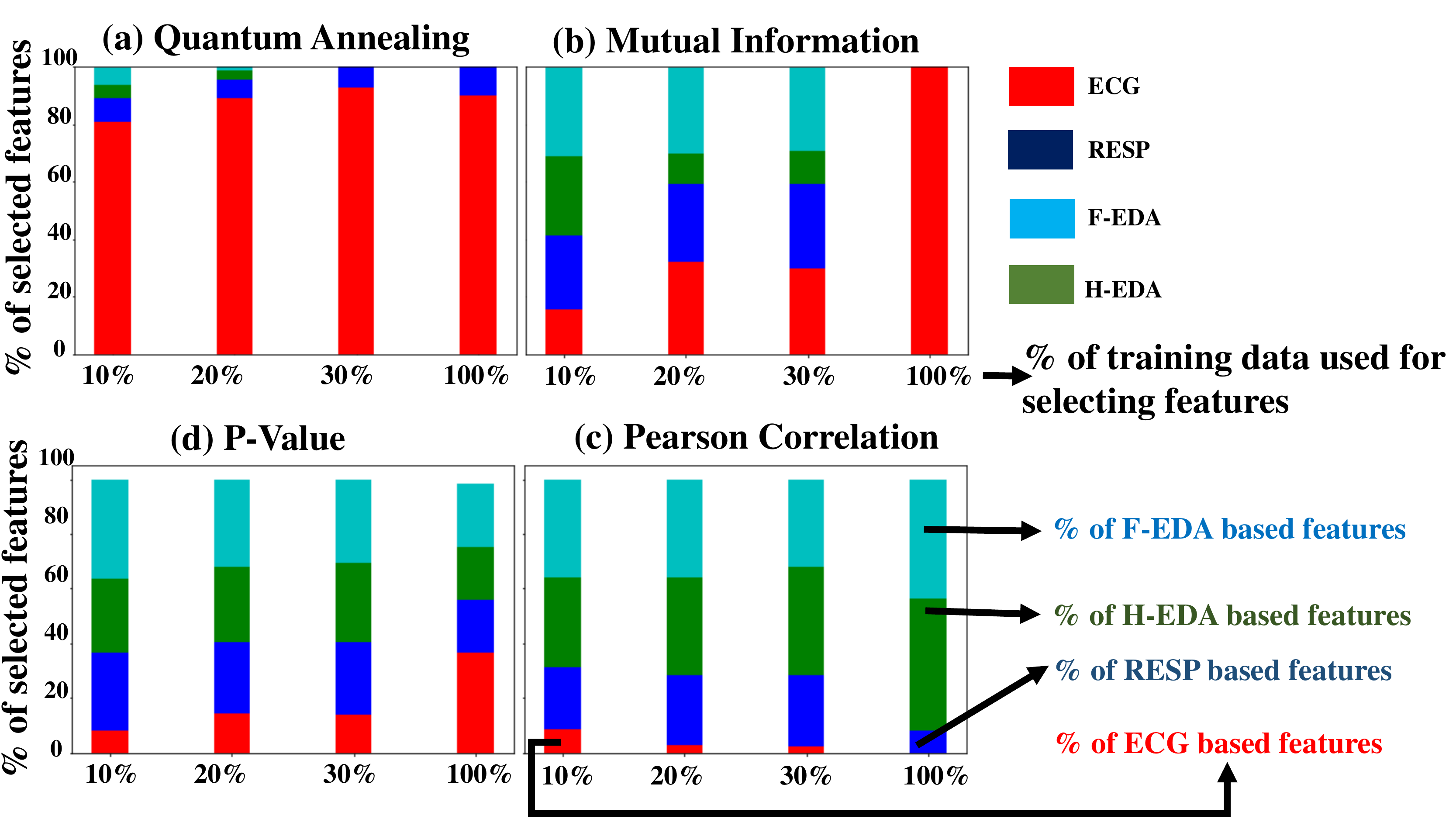}
\caption{Clockwise from top: (a) quantum annealing based selection, (b) mutual information based selection, (c) Pearson correlation based feature selection, and (d) p-value based feature selection.}
\label{fig_4_label}
\end{figure*}

\subsection{Qualitative Evaluation}
In this section, we will perform a qualitative evaluation of the features selected using four techniques namely quantum annealing, mutual information, p-value, and Pearson correlation feature selection technique. These four techniques were used to extract features using 100\%, 30\%, 20\%, and 10\% of the training data. For each of the techniques, the feature selection procedure was repeated n-times (n=10 in our case) to provide less importance to features that got selected just by random chance. The selected features were categorized according to their sources such as ECG-based features, H-EDA-based features, F-EDA-based features, and RESP-based features. Figure \ref{fig_4_label} shows the contribution of specific signal sources in terms of the percentage of total features selected.

From Figure \ref{fig_4_label}, we can visualize that quantum annealing based feature selection returned a more stable set of features from mainly two signal sources (ECG and RESP) for 100\% training data and 30\% training data. Even with 20\% and 10\% of the training data, ECG and respiration-based features contributed to more than 95\% and 88\% of the total feature count. This shows that QA-based feature selection consistently found EDA-based features less important than ECG and RESP-based features. This consistency is not affected when only 30\% of the training data is available for feature selection. This consistency deteriorates only slightly even when 10\% and 20\% of the training data are available.

\subsection{Quantitative Evaluation}
\begin{figure*}[h]
\centering
\includegraphics[trim= 0cm 0cm 0cm 0cm, scale=0.42]{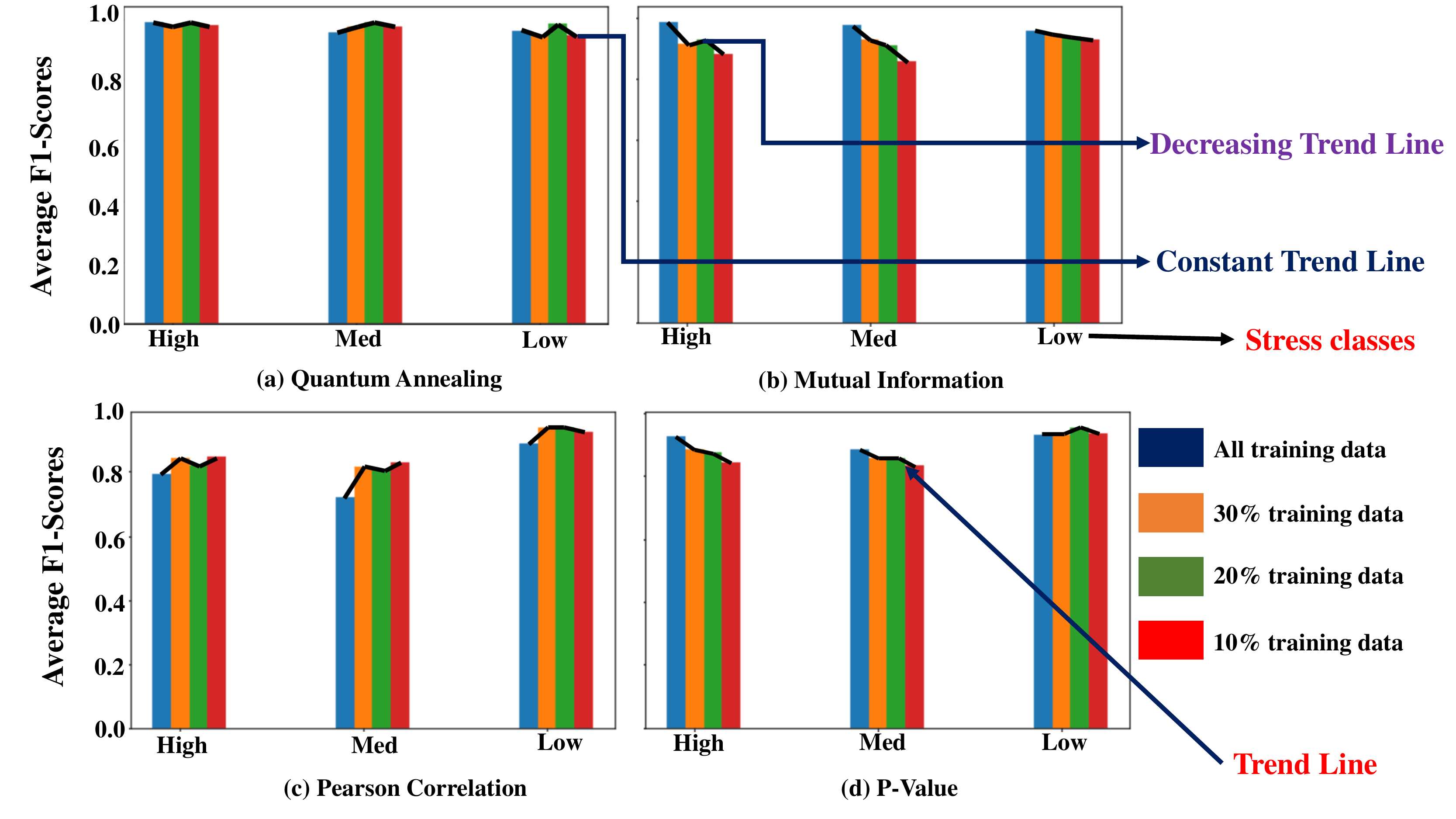}
\caption{Clockwise from top: (a) quantum annealing based selection, (b) mutual information based selection, (c) p-value based feature selection, and (d) pearson correlation based feature selection . Trend line is shown in black and is obtained by connecting the F1-scores obtained under different training sizes for high, medium, and low stress.}
\label{fig_6_label}
\end{figure*}

The quantitative evaluation will be based on the average F1-score computed under all training data scenarios. The training data scenarios are 100\% training data, 30\% training data, 20\% training data, and 10\% training data. The effectiveness of a particular feature selection algorithm will be estimated by the ability of the selected features to classify a feature sample in low, medium, and high stress classes using classical classification. The metric used to quantify the classification performance is the F1-score. Figure \ref{fig_6_label} shows the average F1-score achieved when features were selected using the different training data sizes for QA-based feature selection technique, mutual-information-based feature selection, Pearson-correlation-based feature selection, and p-value-based feature selection.

From Figure \ref{fig_6_label}, it can be visualized that the F1-score for quantum-annealing-based feature selection technique is more or less constant when features are selected using different training sizes for high, medium, and low-stress classes. The trend line of the F1-score obtained using different training sizes does not follow a clear increasing or decreasing trend. This implies that the average F1-score did not drop significantly as training samples were reduced. Moreover, no significant statistical difference (p-value$>$0.05) was observed between the F1-scores  obtained under 30\%, 20\%, and 10\% training data and the F1-score obtained when 100\% of the training data was used for feature selection. This implies that the performance of QA-based feature selection was not affected to a significant extent under the presence of limited training data when the quantum-annealing-based feature selection technique was used.

In the case of the mutual-information-based feature selection technique, the trend line follows a decreasing trend as training sizes are reduced. Moreover, except for low stress, a significant statistical difference (p-value$<$0.05) is observed between the  F1-scores  obtained under 30\%, 20\%, and 10\% training data and the F1-score obtained when 100\% of the training data was used. Similarly, for the p-value-based feature selection technique, a clear decreasing trend is observed with the average F1-score as the size of training data is reduced. An exception to this observation is the situation where p-value-based features were used to detect low stress. For the p-value-based feature selection technique, no significant statistical difference was observed between the F1-scores obtained under 30\%, 20\%, and 10\% training data and the F1-score obtained when 100\% of the training data. However, the average F1-score obtained using the p-value-based feature selection technique is significantly lower than that obtained using quantum annealing and mutual-information-based feature selection technique. Pearson-correlation-based feature selection technique did not perform as well as compared to other feature selection techniques which are evident from the overall low average F1-score across all groups and training data size. 

\section{Conclusion and Future Work}
\label{conc}
In this paper, a novel method for automated feature selection using quantum annealing is explored. We have proposed a hybrid classical-quantum machine learning pipeline that integrates D-Wave's QPU through a software interface that allows interaction with the physical QPU. The problem of selecting optimal features is formulated as selecting a clique with the lowest energy from a binary quadratic model. The nodes forming the lowest energy clique are returned as the set of optimal feature space. As a case study, QA-based feature selection was used to select the optimum feature subset for detecting stress in a driving scenario and compared with mutual information, Pearson coefficient, and p-value based feature selection techniques. Results show that QA based feature selection technique performed equally well as the classical techniques. In conditions where training data was limited, a slight advantage is observed when QA based feature selection algorithm was used. However, future investigation on large-scale datasets and/Or on different application domain will help us understand how quantum-annealing-based optimization techniques generally performs better than the classical techniques.

Preliminary results from this study shows the promise of quantum annealing as an efficient alternative to classical feature selection techniques. QA-based feature optimization technique was found to return consistent results both in terms of performance and the consistency of features selected even when training sizes were reduced in size. In the future, we plan to extend our analysis of quantum-annealing based feature selection techniques in solving diverse problems where feature selection is an important step in the machine learning pipeline.

\bibliographystyle{IEEEtran}
\bibliography{references}

\end{document}